\documentclass{article}
\usepackage{spconf,amsmath,graphicx,url,amssymb}
\usepackage{subfigure}
\usepackage{graphicx}
\usepackage{color}
\usepackage{enumitem}
\usepackage{units}
\usepackage{listings}
\usepackage{booktabs}

\usepackage{courier}

\definecolor{olive}{rgb}{0.63, 0.62, 0.25}

\title{To catch a chorus, verse, intro, or anything else: Analyzing a song with structural functions}
\name{Ju-Chiang Wang$^{1}$, Yun-Ning Hung$^{2,}$\sthanks{The author conducted this work as an intern at ByteDance.}, and Jordan B. L. Smith$^{1}$}
\address{$^{1}$ ByteDance \\ 
$^{2}$ Center for Music Technology, Georgia Institute of Technology, Atlanta, GA, USA \\
{\small\tt ju-chiang.wang@bytedance.com, amy.hung@gatech.edu, jordan.smith@bytedance.com}}

\begin{document}
\ninept
\maketitle
\begin{abstract}
Conventional music structure analysis algorithms aim to divide a song into segments and to group them with abstract labels (e.g., `A', `B', and `C'). However, explicitly identifying the function of each segment (e.g., `verse' or `chorus') is rarely attempted, but has many applications. We introduce a multi-task deep learning framework to model these structural semantic labels directly from audio by estimating ``verseness,'' ``chorusness,'' and so forth, as a function of time. We propose a 7-class taxonomy (i.e., intro, verse, chorus, bridge, outro, instrumental, and silence) and provide rules to consolidate annotations from four disparate datasets. We also propose to use a spectral-temporal Transformer-based model, called SpecTNT, which can be trained with an additional connectionist temporal localization (CTL) loss. In cross-dataset evaluations using four public datasets, we demonstrate the effectiveness of the SpecTNT model and CTL loss, and obtain strong results overall: the proposed system outperforms state-of-the-art chorus-detection and boundary-detection methods at detecting choruses and boundaries, respectively.

\end{abstract}


\section{Introduction}\label{sec:introduction}


In Music Structure Analysis (MSA) of audio, two tasks are defined: \emph{segmentation}, where the aim is to divide a recording of a song into non-overlapping segments, and \emph{labeling} (or grouping), where the aim is to label the segments with symbols (e.g., `A', `B', etc.) to indicate how they are grouped. However, these bits of information give a limited description of a song. More advanced tasks include \emph{hierarchical analysis}, in which segmentation and grouping must be estimated at multiple timescales, and \emph{semantic labeling}, in which the labels should have meaning, such as `chorus', `verse', or `solo'. 
While hierarchical MSA has attracted sustained attention~\cite{mcfee2014analyzing, mcfee2015hierarchical, grill2015music, tralie2019enhanced, salamon2021deep-embeddings}, semantic MSA, despite having many uses, has rarely been addressed: the most significant work on this task was over a decade ago \cite{paulus2009signal, paulus2010labelling}.

%
Early MSA algorithms (see~\cite{nieto2020audio} for a review) were mostly non-supervised,
based on processing some version of a self-similarity matrix computed from traditional audio features such as MFCCs or chroma. However, as large annotated datasets became available (e.g., \emph{SALAMI}~\cite{smith2011design} and \emph{Harmonix Set}~\cite{nieto2019harmonix}), supervised approaches benefitted.
For example, the usual way to estimate segment boundaries was once to compute a `novelty function' and choose the points that maximized it~\cite{foote2000automatic}; Ullrich et al. used a subset of~\cite{smith2011design} to train a model to estimate the `boundaryness' of each instant given a long context window, and achieved a new state-of-the-art~\cite{ullrich2014_ismir}.
Similarly, Wang et al.~\cite{wang2021supervised} trained a model to estimate `chorusness' (jointly with boundaries), and achieved state-of-the-art \emph{chorus detection}. (This is another sub-task of MSA, and an early work on this topic~\cite{bartsch2001catch} is alluded to by our title.) In this paper, we extend~\cite{wang2021supervised} and aim to model `verseness', `bridgeness', and more, directly from audio.
Our system represents the first that is able to assign generic structural semantic labels since~\cite{paulus2009signal}, and the first in which the audio content directly informs the label prediction.


We build on~\cite{wang2021supervised} by introducing two deep neural network (DNN) architectures to MSA, \emph{Harmonic-CNN} \cite{won2020data} and \emph{SpecTNT} (Spectral-Temporal Transformer in Transformer) \cite{specTNT}.
Transformer architectures have not been used yet for MSA, but have the potential to improve the temporal modeling at the representation learning stage; they have earned great attention for their strength in modeling sequential data~\cite{vaswani2017attention,devlin2018bert}.
In particular, we adopt SpecTNT, a hierarchical variant of Transformer that models the spectral and temporal dimensions of an input spectrogram with two Transformer encoders.
We also propose to use the Connectionist Temporal Localization (CTL) loss \cite{wang2019connectionist} to improve the temporal modeling.


A system capable of estimating semantic labels has clear uses: it would enhance any application that already relies on MSA, such as structure-based navigation systems~\cite{goto2003smartmusickiosk} or
automated remixing of sections from different songs~\cite{davies2014automashupper,huang2021modeling}.
Furthermore, the ability to analyze structure without requiring full context of a song allows MSA to be applied to song fragments, or songs that are very short (e.g., a TikTok song under 30 seconds long). For such inputs, approaches that rely on self-similarity matrices and clustering may fail.
An automatic mastering system~\cite{deman2019intelligentmusic} could be improved by predicting the function and structure of the input clips.
Finally, a system not dependent on a full song would be a first step toward a \emph{real-time} MSA system. Such a system could be essential to applications in live contexts, such as a system that controlled the visuals at a concert to match the music (as suggested in 
\cite{nieto2020audio}).

To sum up, our proposed MSA system differs from most previous work in three ways: it predicts semantic (not abstract) labels; it is supervised; and it is non-context-dependent, meaning it does not rely on detecting repeating sequences or relative novelty changes.
The main contributions of this paper are: a mapping used to consolidate information from disparate datasets (Sec. 2); the proposed approaches, including a system that represents the first application of Transformers to MSA (Sec. 3); and the evaluations, including ablation and cross-dataset studies using four public datasets (Sec. 4). 


\section{Structural Label Conversion}
\label{sec:anno_map}

%

Structural analysis is recognized as an ``ill-defined'' problem~\cite{nieto2020audio}, since the ``solution'' is not unique: listeners usually disagree about the exact structure of a piece.
The SALAMI dataset was designed to mitigate this: section groupings were annotated separately from function (inspired by~\cite{peeters2009is-music}), and an ``Annotator's Guide''~\cite{salami_github}
defined a valid set of function labels.
The data still contain many instances of conflicting labels, including dozens of cases where one listener's `chorus' was another's `verse'.
However, there is far more agreement than disagreement:
if two annotators agree on the start time of a section, they agree on the function label at least twice as often as they disagree,
and this rate only increases if we collapse synonymous groups of labels: e.g., the Annotator's Guide indicates `coda', `fade-out,' and `outro' all have the same basic function.
Therefore, we expect the data created by annotators to be useful for this task.

To train a model to predict functions, we must strike a balance between our goal---to model a rich set of semantic meanings---with our available means: a motley set of datasets, collected with different standards, made up of different genres, with functional terms that vary in specificity.
Based on our study of existing datasets, there is no standard naming system for structural labels.
For example, some use `refrain' as chorus, some specify section sub-types (e.g., `verse A' in RWC~\cite{goto2002rwc}), and some allow compound functions (e.g., `\mbox{instchorus}', appearing in Harmonix Set). 
Before developing a model to classify structural functions, we have to define a fixed taxonomy, and then define rules to map the free-form annotations from several datasets onto the same taxonomy.
Any choice here will be a compromise among the utility of the taxonomy, the amount of training data per class, and the validity of the mappings.





We chose a 7-class taxonomy that is suited to Western pop music: `intro', `verse', `chorus', `bridge', `inst' (i.e., instrumental), `outro', and an auxiliary class `silence'.
This is equivalent to the set of 5 basic classes named in the SALAMI Annotator's Guide, plus `inst' and `silence'.
We then define Algorithm 1, which gives a mapping for any annotated label that does not match the 7-class taxonomy. The function \texttt{\footnotesize conversion()} iteratively tests for the existence of substrings in the label. Since multiple substrings can appear in one label, the order of elements in \texttt{\footnotesize substrings} determines the priority of the matches; for instance, `instrumentalverse' will be converted to `verse,' not `inst', and `pre-chorus' will be converted to `verse' to disentangle the build from the true chorus (as in~\cite{wang2021supervised}).
Note that \texttt{\footnotesize "end"} is used to mark the timestamp for the end of a song, and it is not regarded as a function label.
We found that \texttt{\footnotesize substrings} covers 99.3\% of raw labels in the existing datasets. 
The remaining labels mostly include instrument descriptors (such as `guitar,' `gtr,' `riff,' `spoken,' `voice,' and `groove'); after listening to most examples, we decided to convert them all to \texttt{\footnotesize "inst"}.

\begin{figure}[t]
\lstset{language=Python}
\lstset{frame=lines}
\lstset{caption={The conversion rule for a raw label.}, captionpos=b}
\lstset{label={lst:code_direct}}
\lstset{basicstyle=\scriptsize\ttfamily,
        breaklines=true,
        breakatwhitespace=true}
\begin{lstlisting}
substrings = [
   ("silence", "silence"), ("pre-chorus", "verse"), 
   ("prechorus", "verse"), ("refrain", "chorus"), 
   ("chorus", "chorus"),   ("theme", "chorus"), 
   ("stutter", "chorus"),  ("verse", "verse"), 
   ("rap", "verse"),       ("section", "verse"), 
   ("slow", "verse"),      ("build", "verse"), 
   ("dialog", "verse"),    ("intro", "intro"), 
   ("fadein", "intro"),    ("opening", "intro"),  
   ("bridge", "bridge"),   ("trans", "bridge"), 
   ("out", "outro"),       ("coda", "outro"), 
   ("ending", "outro"),    ("break", "inst"), 
   ("inst", "inst"),       ("interlude", "inst"),
   ("impro", "inst"),      ("solo", "inst")]

def conversion(label):
    if label == "end":  return "end"
    for s1, s2 in substrings:
        if s1 in label.lower():  return s2
    return "inst"
\end{lstlisting}
\end{figure}




\section{Proposed Approach}
\label{sec:method}

\begin{figure}[t]
  \centering
  \includegraphics[width=.95\columnwidth, height=11mm]{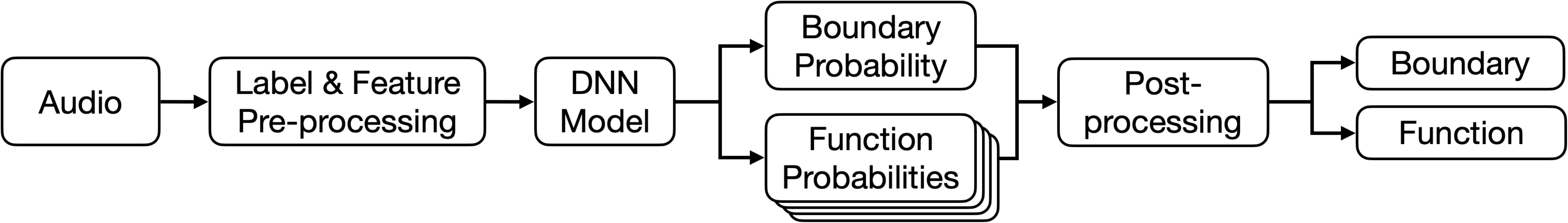}
  \caption{Pipeline for the proposed system.}
  \label{fig:pipeline}
\end{figure}

Our proposed system pipeline is depicted in Fig.~\ref{fig:pipeline}, and is conceptually similar to the pipeline proposed in \cite{wang2021supervised}.
A song is first divided into overlapping audio chunks. After audio feature extraction, the DNN model predicts, for each chunk, two types of activation curves: one for boundaries and one for each structural function.
We propose two alternative strategies, \emph{instant} and \emph{multi-point}, for the DNN model.
In the \emph{instant} model,
each chunk leads to a single prediction vector indicating the likelihood of each function label for the instant at the center of the chunk (as in~\cite{ullrich2014_ismir}).
Then, the sequence of predictions forms the output curves.
In the \emph{multi-point} model, predictions are made for every time point of an input chunk. We follow the method of~\cite{wang2021supervised} to merge the outputs of all the overlapping chunks into the final prediction curves.
Finally, a few simple post-processing steps return the output is a usable format: boundary timestamps and a single function label per segment.


\subsection{Label and Audio Feature Pre-processing}
\label{sec:fea_label}

The steps to create a \emph{chorus activation curve}, the target curve for the prediction of chorus segments, are given in~\cite{wang2021supervised}. We use the same steps, but repeat them for the 6 additional function classes to create 7 \emph{function activation curves}.
There is still only one \emph{boundary activation curve}, but it now includes all boundaries (not just those between chorus and non-chorus segments), as described in~\cite{ullrich2014_ismir}.

We smooth the transitions of the function activation curves using a 2-second-wide Hann window: a 1-second ramp from 0 to 1 prior to the onset, and a 1-second ramp down after the offset, as in~\cite{wang2021supervised}.
Regarding the ``boundary section,'' we set a duration of 0.6 seconds for each boundary (whereas \cite{wang2021supervised} used 0.5 seconds).

As for audio feature extraction, we use harmonic representation (whereas \cite{wang2021supervised} used mel-spectrogram), since trainable harmonic filters help capture harmonic information while preserving spectral-temporal locality.
It has proven useful in music auto-tagging \cite{won2020data}.



\subsection{Harmonic-CNN}

Harmonic-CNN \cite{won2020data} is designed to mimic human perception by extracting the spectrogram features through its harmonic representation front-end, as harmonic structure is known to play a key role in the human auditory system. 
Its network architecture contains several temporal pooling layers with the aim of predicting a single target for each music tag, so we believe it can be a good fit to the goal of the \emph{instant} model in this work.
Given the harmonic representation input, it applies seven 2D-convolutional layers in the back-end to extract high-level features, followed by two dense layers to predict the target of a time-step for the boundary and function activations.


\subsection{Spectral-Temporal Transformer}

Although the Transformer architecture has shown remarkable performance in modeling sequential data such as text \cite{devlin2018bert} and musical scores \cite{huang2018music,zeng2021musicbert}, its data-hungry nature makes it inapt for tasks with little training data. The total number of current annotated songs for public MSA datasets is less than 3000, which is likely insufficient to train a satisfactory Transformer model. 

By contrast, SpecTNT (adopted in this work) has shown good performance in beat and downbeat tracking \cite{hung2022modeling}, vocal melody extraction \cite{specTNT}, and chord recognition \cite{specTNT} using datasets even smaller than those for MSA.
The basic principle of SpecTNT arises from the interaction between two levels of Transformer encoders, namely a \emph{spectral encoder} and a \emph{temporal encoder}. The spectral encoder is responsible for extracting the spectral features via \emph{Frequency Class Tokens} (FCTs) for each time-step, where an FCT is an aggregated embedding that characterizes harmonic and timbral information. 
The temporal encoder then exchanges local information (i.e., FCTs) along the time axis; this self-attention step can help discover structural patterns related to novelty, homogeneity, and repetition~\cite{nieto2020audio}.
For example, the self-attention mechanism can allow frames around a boundary to attend to the boundary, and frames with the same function to attend to one another.
Owing to its hierarchical design, SpecTNT permits a smaller number of parameters as compared to the original Transformer~\cite{vaswani2017attention}, and we expect this attribute can help improve the generalization for the MSA task.



A SpecTNT model consists of three modules: a two-dimensional residual network (ResNet)~\cite{Won2020EvaluationOC} at the front-end to extract intermediate information from the input harmonic representation; a stack of SpecTNT blocks; and a linear layer to output the target probability at all time-steps.
Therefore, it serves as the \emph{multi-point} model in this work.
For ResNet, each convolutional layer uses a kernel size of 3. Then, we use 5 SpecTNT blocks. In each block, we use 96 feature maps with 4 attention heads for the spectral encoder, and 96 feature maps with 8 attention heads for the temporal encoder. 


\begin{figure}[t]
  \centering
  \includegraphics[width=\columnwidth]{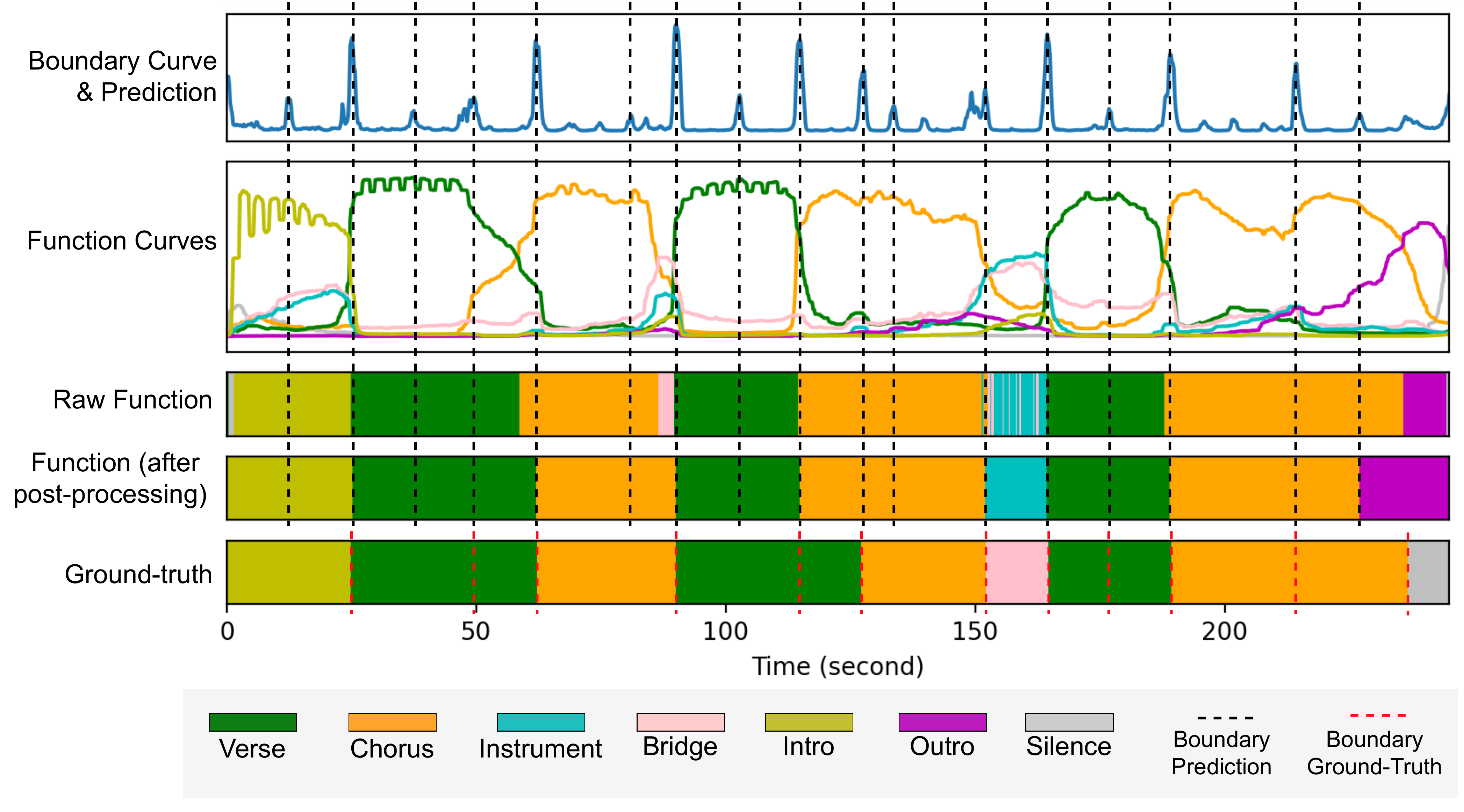}
  \caption{Example of post-processing for a test song \emph{Complicated} by Avril Lavigne, using SpecTNT (with 24s and CTL). The top two rows display the raw outputs of the activation curves. ``Raw Function'' shows the argmax labels at each time-step
  of the DNN output.
  }
  \label{fig:post-process}
\end{figure}

\subsection{Post-processing} 

The raw outputs include a boundary curve and 7 structural function curves.
In an end-to-end manner, one can assign the structural function with the largest probability (i.e., arg-max) for each time-step. However, as illustrated in Fig. \ref{fig:post-process},
this `raw function' output can be inaccurate near boundaries and inconsistent within a segment.
To get more usable output, we first interpret the boundary activation curve using the peak-picking method proposed in~\cite{ullrich2014_ismir}.
Then, we simply choose the function labels with the largest \textit{average} probability in each segment.
After this post-processing, we can see a more accurate result
(see Fig. \ref{fig:post-process}).

\subsection{Training Loss}

To jointly model the boundaries and structural functions, we define two types of losses: \emph{boundary loss} and \emph{function loss}.
Each is calculated by summing the weighted binary cross-entropy between the prediction and target at each time-step of the corresponding activation curve. Following the heuristic used in \cite{wang2021supervised}, we use a weight 0.9 for boundary loss and 0.1 for function loss when combining them, since the boundary curve is sparse and more difficult to learn.

To enhance the coherence of the predictions and reduce the fragmentation,
we propose to add the \emph{Connectionist Temporal Localization (CTL)} loss \cite{wang2019connectionist} to the function loss.
The CTL loss aims to model the sequential order of labels.
In popular music, the order of sections follows regular patterns: a song is unlikely to start with an outro or end with an intro---this is the premise of~\cite{paulus2010labelling}.
We use the raw section annotations (i.e., before they are converted to the activation curves) to form the target sequence of section tokens (e.g., [`intro', `verse', `chorus', `inst']). Then, the CTL loss is calculated between the prediction (a $T \times 7$ probability matrix) and the sequence of $S$ tokens, where $T$ and $S$ are the numbers of time-steps and section tokens, respectively, in a training chunk.

\section{Experiments}\label{sec:experiment}

\subsection{Implementation Details}

For STFT before the harmonic representation, we adopt a window length of 1024 with a hop size of 512 on audio signals of 16 kHz sampling rate.
The time resolution for the activation curves is about 5.2 per second (i.e., a target per 0.192 seconds). We found using longer duration for the input chunks led to better function prediction, but poorer boundary detection accuracy. Thus, we adopt 24-second chunks and will show result of using 36 seconds in an ablation study.


Data augmentation is vital to training an effective Transformer model.
We use the \texttt{\footnotesize torchaudio\_augmentations} package \cite{spijkervet_torchaudio_augmentations} to randomly add noise, adjust gain, or apply high/low-pass filters for each training sample. 
We use a random sampling strategy to load a random chunk into a batch, instead of sequentially loading a chunk from the beginning of a training song as adopted in \cite{wang2021supervised}.
That is, we first enumerate all the valid training chunks into a list by using a 24-second sliding window with a 3-second hop on every song. If the last chunk exceeds the end of a song, we pad zeros at its end. Then, the data loader draws a chunk uniformly from the list during training. Therefore, a batch can include chunks from different locations of different songs. This technique can lead to faster convergence and better results compared to that in \cite{wang2021supervised}.


We use PyTorch 1.8 and Adam optimizer \cite{kingma2014adam} with 0.0005 learning rate, 0.9 weight decay, and 2 epochs of patience. An epoch runs 500 mini-batches with a batch size of 128. We conduct 100 training epochs and use the model with the best validation result for testing. 

\subsection{Experimental Configurations}

We use four public datasets for experiments:
\emph{Harmonix Set} \cite{nieto2019harmonix}, \emph{SALAMI-pop} \cite{smith2011design}, \emph{RWC-Pop} \cite{goto2002rwc}, and \emph{Isophonics} \cite{mauch2009omras2}. 
\emph{Harmonix Set} contains 912 western pop songs. Since the original audio was not available, we searched for the correct audio versions and manually refined the annotations. For \emph{SALAMI-pop}, we select a subset of 274 songs (with 445 annotations) in the ``popular'' genre. 
In these annotations, the label `no\_function' appears often, but this is the result of a parsing error; we replace every instance of it with the preceding section's function label.
\emph{RWC-Pop} is the subset of 100 popular songs in \emph{RWC}. We use the original annotations by AIST.  \emph{Isophonics} contains 277 songs from The Beatles, Carole King, Michael Jackson, and Queen, but we use the \emph{TUT-Beatles} annotations \cite{paulus2010improving} for the 174 Beatles songs. 

We conduct evaluations of two types. First, in the ablation study, we use \emph{Harmonix Set} in a 4-fold cross-validation (CV) manner, but include \emph{SALAMI-pop}, \emph{RWC-Pop}, and \emph{Isophonics} in the training set at every iteration. Second, we carry out cross-dataset evaluations: each of \emph{SALAMI-pop}, \emph{RWC-Pop}, and \emph{Isophonics} in turn serves as the test set, and the remaining datasets are used for training. We randomly split 10\% of the training set to form the validation set.

Besides regular MSA, we are also interested in the performance of chorus detection, where we only leave the boundary and section annotations involving a `chorus.'
The \texttt{\footnotesize mir\_eval} package \cite{raffel2014mir_eval} is used to compute the following evaluation metrics:
(1) \emph{HR.5F}: F-measure of hit rate at 0.5 seconds; (2) Accuracy (\emph{ACC}): the frame-wise accuracy between the predicted function label and the converted ground-truth label; (3) \emph{PWF}: F-measure of pair-wise frame clustering; (4) \emph{Sf}: F-measure of normalized entropy score; (5) \emph{CHR.5F}: F-measure of `chorus' boundary hit rate at 0.5 seconds; (6) \emph{CF1}: F-measure of pair-wise frames for `chorus' and `non-chorus' sections \cite{wang2021supervised}. The definitions of (1), (3), (4), and (5) can be found in \cite{nieto2020audio}.


\subsection{Baseline Methods}
\label{sec:discussion}

We include several baseline methods from prior work. ``Scluster'' is the spectral clustering algorithm \cite{mcfee2014analyzing} implemented in MSAF \cite{nieto2016systematic}. 
``DSF + Scluster'' uses learned deep structure features (DSF) \cite{wang2021deepstruc} as the inputs for Scluster to predict the MSA results. It employs a Harmonic-CNN trained with metric learning loss in a supervised fashion, and is trained only on \emph{Harmonix Set}. 
To perform chorus detection for the above two methods,
we pick chorus sections using the heuristic in \cite{wang2021supervised},
i.e., \emph{Max-dur}, which chooses the segment group that covers the greatest duration of a song as the choruses.
``CNN-Chorus'' \cite{wang2021supervised} uses a similar approach to this work, but with a different DNN model and training strategy. It was developed to only detect the chorus sections. 
For \emph{RWC-Pop} and \emph{Isophonics}, which are used in MIREX, we include for comparison the results of three top-performing MIREX submissions: GS3~\cite{grill2015music}, and SMGA1 and SMGA2~\cite{serra2012importance}.

\section{Discussion and Conclusion}
\label{sec:conclusion}

Results are presented in Table \ref{table:result}. 
The main proposed systems are ``Harmonic-CNN'' (the \textit{instant} method) and ``SpecTNT (24s, CLT)'' with CTL loss (the \textit{multi-point} method). In the ablation study, we compare these to SpecTNT without CTL; a regular Transformer~\cite{won2021transformer} with CTL; and SpecTNT with CTL and a longer chunk size (36s).
The regular Transformer contains only the temporal encoders \cite{devlin2018bert}, so it is non-hierarchical and has no spectral encoder.
It is clear that this system performs worse than SpecTNT, most likely because the training data is insufficient.
Comparing SpecTNT with and without CTL loss, we can see that including the loss improves performance slightly on all metrics, demonstrating its usefulness.
We also observe that SpecTNT with longer inputs can improve the function prediction, but the boundary detection performance drops, possibly because the model has over-fit the training data. To prioritize boundary accuracy, we use the shorter window in the next evaluation.

\begin{table}[t]
\centering
\resizebox{\columnwidth}{!}{
\begin{tabular}{l|cccccc} 
\toprule
 & HR.5F & ACC & PWF & Sf & CHR.5F & CF1  \\
 \midrule
 \multicolumn{7}{c}{\textbf{Ablation Study}} \\
 \midrule
 &  \multicolumn{6}{l}{~~~~\textit{Harmonix Set}} \\
Scluster~\cite{mcfee2014analyzing}  & .263 & - & .586 & .641 & .171 & .534 \\
DSF + Scluster~\cite{wang2021deepstruc} & .497 & - & .689 & \textbf{.743} & .326 & .611 \\
CNN-Chorus~\cite{wang2021supervised} & - & - & - & - & .371 & .692 \\
Harmonic-CNN & .559 & .680 & .670 & .682 & .462 & .784 \\
Transformer (24s, CTL) & .521 & .640 & .655 & .649 & .399 & .755 \\
SpecTNT (24s)     & .565 & .690 & .687 & .702 & .491 & .813 \\
SpecTNT (24s, CTL) & \textbf{.570} & .701 & .700 & .714 & \textbf{.501} & .815 \\
SpecTNT (36s, CTL) & .558 & \textbf{.723} & \textbf{.712} & .724 & .476 & \textbf{.831} \\

 \midrule
 \multicolumn{7}{c}{\textbf{Cross-dataset Evaluation}} \\
 \midrule
 
 &  \multicolumn{6}{l}{~~~~\textit{SALAMI-pop} (subset of MIREX 2012 dataset)}  \\
[0.1cm]
Scluster~\cite{mcfee2014analyzing}  & .305 & - & .545 & .572 & .196 & .418 \\
DSF + Scluster~\cite{wang2021deepstruc}  & .447 & - & .615 & \textbf{.653} & .272 & .573 \\
CNN-Chorus~\cite{wang2021supervised} & - & - & - & - & .308 & .602 \\
Harmonic-CNN   & .477 & .525 & .631 & .636 & .340 & .777 \\
SpecTNT (24s, CTL) & \textbf{.490} & \textbf{.544} & \textbf{.651} & .632 & \textbf{.357} & \textbf{.811} \\
[0.2cm]
 &  \multicolumn{6}{l}{~~~~\textit{RWC-Pop} (MIREX 2010 RWC collection)}  \\
[0.1cm]
GS3 (2015)~\cite{grill2015music} & .524 & - & .542 & .684 & - & - \\
SMGA2 (2012)~\cite{serra2012importance} & .246 & - & .688 & .733 & - & - \\
DSF + Scluster~\cite{wang2021deepstruc} & .438 & - & .704 & \textbf{.739} & .343 & .653\\
Harmonic-CNN   & .571 & .646 & .719 & .694 & .396 & .800 \\
SpecTNT (24s, CTL) & \textbf{.623} & .\textbf{675} & \textbf{.749} & .728 & \textbf{.465} & \textbf{.847} \\
[0.2cm]
 &  \multicolumn{6}{l}{~~~~\textit{Isophonics} (MIREX 2009 Collection)} \\
[0.1cm]
GS3 (2015)~\cite{grill2015music} & .564 & - & .567 & .686 & - & - \\
SMGA1 (2012)~\cite{serra2012importance} & .228 & - & \textbf{.653} & \textbf{.700} & - & - \\
Harmonic-CNN & .543 & .499 & .611 & .598 & .339 & .670\\
SpecTNT (24s, CTL) & \textbf{.590} & \textbf{.550} & .635 & .614 & \textbf{.401} & \textbf{.733} \\

\bottomrule
\end{tabular}}
\caption{Results on four datasets.}
\label{table:result}
\end{table}

Looking at all the studies, we observe that SpecTNT (24s, CTL) consistently outperforms Harmonic-CNN in all metrics; i.e., the \textit{multi-point} model beats the \textit{instant} model.
This may be due to improved temporal coherence in the estimates provided by SpecTNT's hierarchical Transformer.
Either way, the multi-point method used with SpecTNT is 15.6 times more efficient at prediction time than the instant method required by Harmonic-CNN, since the former requires 1 call to the model every 3 seconds (as a chunk hops every 3 sec), whereas the latter requires 5.2 calls per second (i.e., time resolution).
The \textit{multi-point} method also significantly outperforms the Scluster-based methods at boundary detection (i.e., HR.5F and CHR.5F), showing that directly modeling the boundaries with an activation curve is better than doing feature clustering.
In addition, our \textit{multi-point} model can outperform the state-of-the-art boundary detection approach
(i.e., a CNN-based \textit{instant} model, Harmonic-CNN, or GS3) by a wide margin, demonstrating the strength of SpecTNT for boundary detection.
We also find that modeling more function types may help the system to recognize chorusness more accurately: Harmonic-CNN outperforms its predecessor CNN-Chorus (which only saw chorus and non-chorus labels) at chorus detection.

Our proposed approach is less successful in terms of PWF and Sf on \emph{Isophonics}. A closer investigation reveals that there are relatively few `chorus' (or `refrain') labels in the songs of the Beatles, which is a classic evaluation corpus for MSA, but contains many songs with experimental structure. The mismatch between the Beatles songs and the remainder of the datasets, which have more choruses, may account for this poorer performance.
However, inspection of other outputs shows that our model often makes justifiable errors: for example, it often predicts an `outro' when the annotation says `chorus', but when the song is in fact fading out, like a `fade-out chorus'. Fig. \ref{fig:post-process} shows such an example, but also hints at a solution: allowing multiple labels to be potentially predicted per segment.
We have proposed a Transformer-based MSA system that can predict meaningful segment labels, even without context. 
However, in this work we only study the performance of full-track prediction; to demonstrate its robustness to song fragments remains future work.


\bibliographystyle{IEEEbib}
\bibliography{citations}

\end{document}